\documentclass[english,aps,pre,notitlepage]{revtex4}
\usepackage[T1]{fontenc}
\usepackage[latin9]{inputenc}
\usepackage{listings}
\usepackage{babel}
\usepackage{textcomp}
\usepackage{graphicx}
\usepackage[unicode=true]
 {hyperref}

\usepackage{natbib,har2nat}

\makeatletter

\providecommand{\tabularnewline}{\\}

\@ifundefined{textcolor}{}
{%
 \definecolor{BLACK}{gray}{0}
 \definecolor{WHITE}{gray}{1}
 \definecolor{RED}{rgb}{1,0,0}
 \definecolor{GREEN}{rgb}{0,1,0}
 \definecolor{BLUE}{rgb}{0,0,1}
 \definecolor{CYAN}{cmyk}{1,0,0,0}
 \definecolor{MAGENTA}{cmyk}{0,1,0,0}
 \definecolor{YELLOW}{cmyk}{0,0,1,0}
}

\makeatother

\begin{document}

\preprint{This line only printed with preprint option}

\title{Predicting financial markets with Google Trends and not so random
keywords}

\author{Damien Challet}

\email{damien.challet@ecp.fr}

\homepage{http://fiquant.mas.ecp.fr/challet}

\affiliation{Chaire de finance quantitative, Laboratoire de mathématiques appliquées
aux systèmes, École Centrale Paris, Grande Voie des Vignes, 92295
Châtenay-Malabry, France}

\affiliation{Encelade Capital SA, Parc Scientifique C, EPFL, 1015 Lausanne, Switzerland}

\author{Ahmed Bel Hadj Ayed}

\email{ahmed.belhadjayed@ecp.fr}

\affiliation{Chaire de finance quantitative, Laboratoire de mathématiques appliquées
aux systèmes, École Centrale Paris, Grande Voie des Vignes, 92295
Châtenay-Malabry, France}
\begin{abstract}
We discuss the claims that data from Google Trends contain enough
information to predict future financial index returns. We first review the
many subtle (and less subtle) biases that may affect the backtest
of a trading strategy, particularly when based on such data. Expectedly,
the choice of keywords is crucial: by using an industry-grade backtest
system, we verify that random finance-related keywords do not to contain
more exploitable predictive information than random keywords related
to illnesses, classic cars and arcade games. However, other keywords
applied on suitable assets yield robustly profitable strategies, thereby
confirming the intuition of \cite{preis2013quantifyingGT}.
\end{abstract}
\maketitle

\section{Introduction}

Taking the pulse of society with unprecedented frequency and accuracy
is becoming possible thanks to data from various websites. In particular,
data from Google Trends (GT thereafter) report historical search volume
interest (SVI) of given keywords and have been used to predict the
present \cite{choi2012predicting} (called \emph{nowcasting }in \cite{castle2009nowcasting}),
that is, to improve estimate of quantities that are being created
but whose figures are to be revealed at the end of a given period.
They include unemployment, travel and consumer confidence figures
\cite{choi2012predicting}, quarterly company earnings (from searches
about their salient product)s \cite{da2010search}, GDP estimates
\cite{castle2009nowcasting} and influenza epidemics \cite{ginsberg2008detecting}.

Asset prices are determined by traders. Some traders look for, share
and ultimately create information on a variety on websites. Therefore
asset prices should be related to the behavior of website users. This
syllogism has been investigated in details in \cite{da2011search}:
the price returns of the components of the Russell 3000 index are
regressed on many factors, including GT data, and these factors are
averaged over all of the 3000 assets. Interestingly, the authors find
\emph{inter alia} a significant correlation between changes in SVI
and individual investors trading activity. In addition, on average,
variations of SVI are negatively correlated with price returns over
a few weeks during the period studied (i.e, in sample). The need to
average over many stocks is due to the amount of noise in both price
returns and GT data, and to the fact that only a small fraction of
people who search for a given keywords do actually trade later.

\cite{preis2013quantifyingGT}'s claim is much stronger: it states
that future returns of the Dow Jones Industrial Average are negatively
correlated with SVI surprises related to some keywords, hence that
GT data contains enough data to predict financial indices. Several
subtle (and not so subtle) biases prevent their conclusions from being
as forceful as they could be. Using a robust backtest system, we are
able to confirm that GT data can be used to predict future asset price
returns, thereby placing their conclusions on a much more robust footing.

\section{Data and Strategy}

Raw asset prices are well described by suitable random walks that
contain no predictability whatsoever. However, they may be predictable
if one is able to determine a set of conditions using either only
asset returns (see e.g. \cite{MarsiliClust} for conditions based
on asset cross-correlations) or external sources of information. Google
Trends provide normalized time series of number of searches for given
keywords with a weekly time resolution%
\footnote{When requesting data restricted to a given quarter, GT returns daily
data.%
}, denoted by $v_{t}$. \cite{preis2013quantifyingGT} propose the
following trading strategy: defining the previous base-line search
interest as $\bar{v}_{t}=\frac{1}{T}\sum_{t'=t-T}^{t}v_{t'}$, the
SVI surprise is $\delta_{t}=v_{t}-\bar{v}_{t-1}$, and the position
to take on a related asset during week $t+1$ is $s_{t+1}=-\mbox{sign }\delta_{t}$.
Nothing prevents to consider the inverse strategy, but average price
reversion over the next one or two weeks with respect to a change
of SVI was already noticed by other authors \cite{da2011search,dzielinski2012measuring}. 

Instead of trying to predict the Dow Jones Industrial Average index,
we use the time series of SPY, which mirrors the Standard and Poors
500 index. This provides a weak form of cross-validation, the two
time series being highly correlated but not identical. For the same
reason, we compute returns from Monday to Friday close prices instead
of Monday to Monday, which keeps index returns in sync with GT data
(they range from Sundays to Saturdays).

\section{Methodological biases}

Prediction is hard, especially about the future. But prediction about
the future \emph{in} the past is even harder. This applies in particular
to the backtesting of a trading strategy, that is, to the computation
of its virtual gains in the past. It is prone to many kinds of biases
that may significantly alter its reliability, often positively \cite{freeman1992behind,leinweber2007stupid}.
Most of them are due to the regrettable and possibly inevitable tendency
of the future to creep into the past.

\subsection{Tool bias}

This is the most overlooked bias. It explains in part why backtest
performances are often very good in the 80s and 90s, but less impressive
since about 2003, even when one accounts for realistic estimates of
total transaction costs. Finding predictability in old data with modern
tools is indeed easier than it ought to be. Think of applying computationally
cpu- or memory-intensive methods on pre-computer era data. The best
known law of the computational power increase is named after Gordon
Moore, who noticed that the optimal number of transistors in integrated
circuits increases exponentially with time (with a doubling time $\tau\simeq$~2
years) \cite{moore1965cramming}. But other important aspects of computation
have been improving exponentially with time, so far, such as the amount
of computing per unit of energy (Koomey' law, $\tau\simeq1.5$~years
\cite{koomey2011computingefficiency}) or the price of storage (Kryder's
law, $\tau\simeq$~2 years \cite{kryder2009hdd}). Remarkably, these
technological advances are mirrored by the evolution of a minimal
reaction timescale in financial data \cite{hardiman2013critical}.
In addition, the recent ability to summon and unleash almost at once
deluges of massive cloud computing power on large data sets has changed
the ways financial data can be analyzed. It is very hard to account
for this bias. For educational purposes, one can familiarize oneself
with past computer abilities with virtual machines such as \texttt{qemu}
\cite{bellard2005qemu} tuned to emulate the speed and memory of computers
available at a given time for a given sum of money. 

The same kind of bias extends to progresses of statistics and machine
learning literature, and even to the way one understands market dynamics:
using a particular method is likely to give better results before
its publication than, say, one or two years later.  One can stretch
this argument to the historicity of the methods tested on financial
data at any given time because they follow fashions. At any rate,
this is an aspect of backtesting that deserves a more systematic study.

\subsection{Data biases}

Data are biased in two ways. First, when backtesting a strategy that
depends on external signals, one must ask oneself first if the signal
was available at the dates that it contains. GT data was not reliably
available before 6 August 2008, being updated randomly every few months
\cite{wiki:GT}. Backtests at previous dates include an inevitable
part of science fiction, but are still useful to calibrate strategies. 

The second problem is that data is revised, for several reasons. Raw
financial data often contains gross errors (erroneous or missing prices,
volumes, etc.), but this is the data one would have had to use in
the past. Historical data downloaded afterwards has often been partly
cleaned. \cite{Daco} give good advice about high-frequency data cleaning.
Revisions are also very common for macro-economic data. For example,
Gross Domestic Product estimates are revised several times before
the definitive figure is reached (about revision predictability, see
e.g. \cite{faust2005news}). 

More perversely, data revision includes format changes: the type of
data that GT returns was tweaked at the end of 2012. It used to be
made of real numbers whose normalization was not completely transparent;
it also gave uncertainties on these numbers. Quite consistently, the
numbers themselves would change within the given error bars every
time one would download data for the same keyword. Nowadays, GT returns
integer numbers between 0 and 100, 100 being the maximum of the time-series
and 0 its minimum; small changes of GT data are therefore hidden by
the rounding process; error bars are no more available, but it is
fair to assume that a fluctuation of $\pm1$ should be considered
irrelevant. In passing, the process of rounding final decimals of
prices sometimes introduces spurious predictability, which is well
known for FX data \cite{NeilPrivateComm}.

Revised data also concerns the investible universe. Freely available
historical data does not include deceased stocks. This is a real problem
as assets come and go at a rather steady rate: today's set of investible
assets is not the same as last week's. Accordingly, components of
indices also change. Analyzing the behavior of the components of today's
index components in the past is a common way to force feed it with
future information and has therefore an official name: survivor(ship)
bias. This is a real problem known to bias considerably measures of
average performance. For instance \cite{freeman1992behind} shows
that it causes an overestimation of backtest performance in 90\% of
the cases of long-only portfolios in a well chosen period. This is
coherent since by definition, companies that have survived have done
well. Early concerns were about the performance of mutual funds, and
various methods have been devised to estimate the strength of this
bias given the survival fraction of funds \cite{brown1992survivorship,elton1996survivor}

Finally, one must mention that backtesting strategies on untradable
indices, such as the Nasdaq Composite Index, is not a wise idea since
no one could even try to remove predictability from them.

\subsection{Choice of keywords}

What keywords to choose is of course a crucial ingredient when using
GT for prediction. It seems natural to think that keywords related
to finance are more likely to be related to financial indices, hence,
to be more predictive. Accordingly, \cite{preis2013quantifyingGT}
build a keyword list from the Financial Times, a financial journal,
aiming at biasing the keyword set. But this bias needs to be controlled
with a set of random keywords unrelated to finance, which was neglected.

Imagine indeed that some word related to finance was the most relevant
in the in-sample window. Our brain is hardwired to find a story that
justifies this apparent good performance. Statistics is not: to test
that the average performance of a trading strategy is different from
zero, one uses a T test, whose result will be called t-stat in the
following, and is defined as $z=\frac{\mu}{\sigma}\sqrt{N}$ where
$\mu$ stands for the average of strategy returns, $\sigma$ their
standard deviation and $N$ is the number of returns; for $N>20$,
$z$ looks very much like a Gaussian variable with zero average and
unit variance. \cite{preis2013quantifyingGT} wisely compute t-stats:
the best keyword, \texttt{debt}, has a t-stat of 2.3. The second best
keyword is \texttt{color} and has a t-stat of 2.2. Both figures are
statistically indistinguishable, but \texttt{debt} is commented upon
in the paper and in the press\texttt{; color} is not, despite having
equivalent ``predictive'' power. 

Let us now play with random keywords that were known before the start
of the backtest period (2004). We collected GT data for 200 common
medical conditions/ailments/illnesses, 100 classic cars and 100 all-time
best arcade games (reported in appendix A) and applied the strategy
described above with $k=10$ instead of $k=5$. Table \ref{tab:tstats}
reports the t-stats of the best 3 positive and negative performance
(which can be made positive by inverting the prescription of the strategy)
for each set of keywords.

\begin{table}
\begin{tabular}{|c|c|c|c|c|c|c|c|c|c|c|}
\hline 
keyword & t-stat &  & keyword & t-stat &  & keyword & t-stat &  & keyword & t-tstat\tabularnewline
\hline 
\hline 
\texttt{multiple sclerosis} & -2.1 &  & \texttt{Chevrolet Impala} & -1.9 &  & \texttt{Moon Buggy} & -2.1 &  & \texttt{labor} & -1.5\tabularnewline
\hline 
\texttt{muscle cramps } & -1.9 &  & \texttt{Triumph 2000} & -1.9 &  & \texttt{Bubbles} & -2.0 &  & \texttt{housing} & -1.2\tabularnewline
\hline 
\texttt{premenstrual syndrome } & -1.8 &  & \texttt{Jaguar E-type} & -1.7 &  & \texttt{Rampage} & -1.7 &  & \texttt{success} & -1.2\tabularnewline
\hline 
\texttt{alopecia} & 2.2 &  & \texttt{Iso Grifo} & 1.7 &  & \texttt{Street Fighter} & 2.3 &  & \texttt{bonds} & 1.9\tabularnewline
\hline 
\texttt{gout} & 2.2 &  & \texttt{Alfa Romeo Spider} & 1.7 &  & \texttt{Crystal Castles} & 2.4 &  & \texttt{Nasdaq} & 2.0\tabularnewline
\hline 
\texttt{bone cancer} & 2.4 &  & \texttt{Shelby GT 500} & 2.4 &  & \texttt{Moon Patrol} & 2.7 &  & \texttt{investment} & 2.0\tabularnewline
\hline 
\end{tabular}

\caption{Keywords and associated t-stats of the performance of a simple strategy
using Google Trends time series to predict \texttt{SPY} \label{tab:tstats}
from Monday close to Friday close prices.}
\end{table}

We leave the reader pondering about what (s)he would have concluded
if \texttt{bone cancer} or \texttt{Moon Patrol} be more finance-related.
This table also illustrates that the best t-stats reported in \cite{preis2013quantifyingGT}
are not significantly different from what one would obtains by chance:
the t-stats reported here being a mostly equivalent to Gaussian variables,
one expects 5\% of their absolute values to be larger that 1.95, which
explains why keywords such \texttt{color} as have also a good t-stat.
Finally, \texttt{debt} is not among the three best keywords when applied
to SPY from Monday to Friday: its performance is unremarkable and
unstable, as shown in more details below.

Nevertheless, their reported t-stats of financial-related terms is
biased towards positive values, which is compatible with the reversal
observed in \cite{da2011search,dzielinski2012measuring}, and with
results of Table 1. This may show that the proposed strategy is able
to extract some amount of the possibly weak information contained
in GT data.

\subsection{Coding errors}

An other explanation for this bias could have been coding errors (it
is not). Time series prediction is easy when one mistakenly uses future
data as current data in a program, e.g. by shifting incorrectly time
series; we give the used code in appendix. A very simple and effective
way of avoiding this problem is to replace all alternatively price
returns and external data (GT here) by random time series. If backtests
persist in giving positive performance, there are bugs somewhere.

\subsection{No out-of-sample}

The aim of \cite{preis2013quantifyingGT} was probably not to provide
us with a profitable trading strategy, but to attempt to illustrate
the relationship between collective searches and future financial
returns. It is however striking that no in- and out-sample periods
are considered (this is surprisingly but decreasingly common in the
literature). We therefore cannot assess the trading performance of
the proposed strategy, which can only be judged by its robustness
and consistency out-of-sample, or, equivalently, of both the information
content and viability of the strategy. We refer the reader to \cite{leinweber2007stupid}
for an entertaining account of the importance of in- and out-of-sample
periods.

\subsection{Keywords from the future}

\cite{preis2013quantifyingGT} use keywords that have been taken from
the editions of the FT dated from August 2004 to June 2011, determined
ex post. This means that keywords from 2011 editions are used to backtest
returns in e.g. 2004. Therefore, the set of keywords injects information
about the future into the past. A more robust solution would have
been to use editions of the FT available at or before the time at
which the performance evaluation took place. This is why we considered
sets of keywords known before 2004.

\subsection{Parameter tuning/data snooping}

Each set of parameters, which include keywords, defines one or more
trading strategies. Trying to optimize parameters or keywords is called
data snooping and is bound to lead to unsatisfactory out of sample
performance. When backtest results are presented, it is often impossible
for the reader to know if the results suffer from data snooping. A
simple remedy is not to touch a fraction of historical data when testing
strategies and then using it to assess the consistence of performance
(cross-validation) \cite{freeman1992behind}. More sophisticated remedies
include White's reality check \cite{white2000reality} (see e.g. \cite{sullivan1999data}
for an application of this method). Data snooping is equivalent as
having no out-of-sample, even when backtests are properly done with
sliding in- and out-of-sample periods.

\begin{figure}
\includegraphics[width=0.4\textwidth]{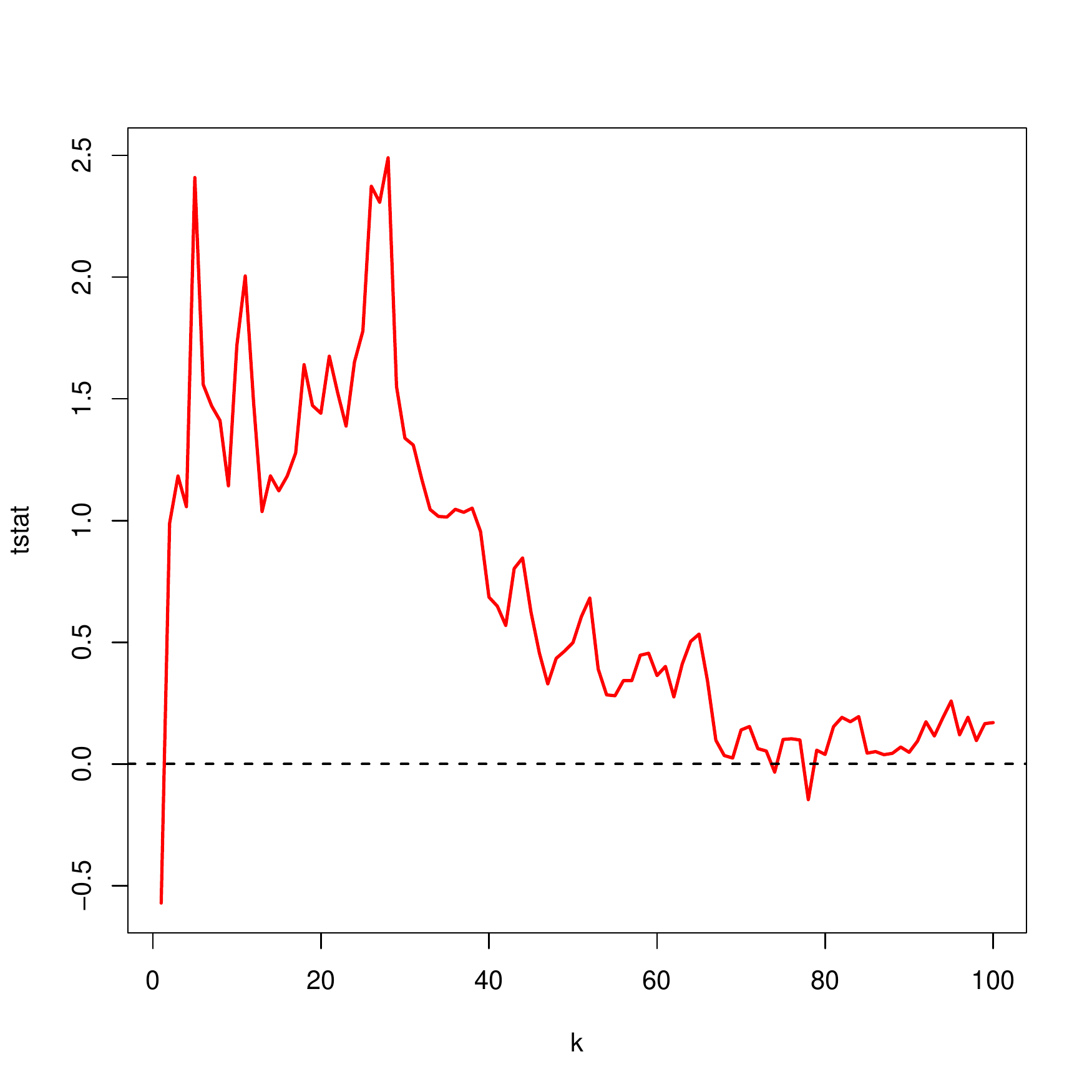}\includegraphics[width=0.4\textwidth]{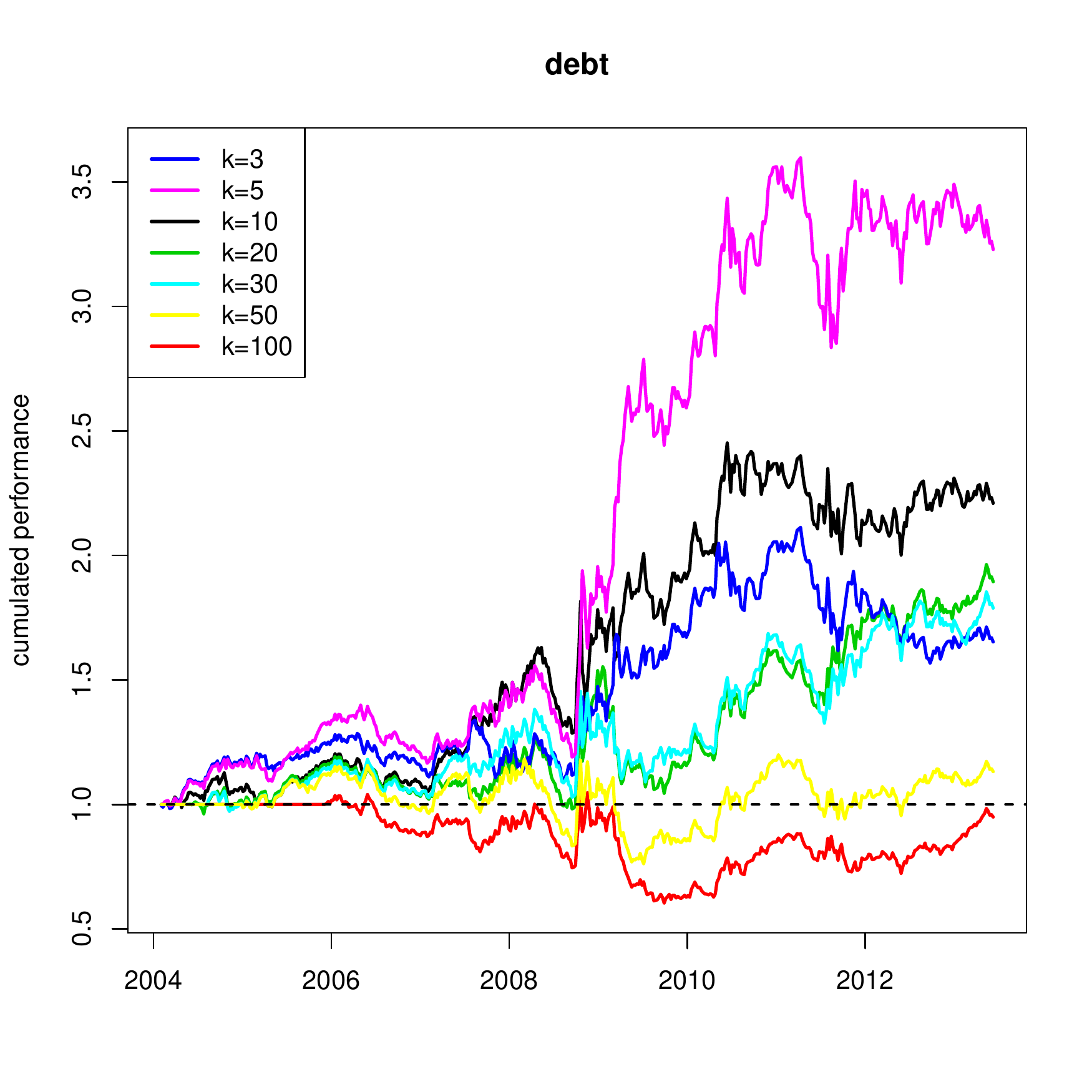}\caption{Left plot: t-stat as a function of the length of the moving average
$k$. Right plot: cumulated performance for various values of $k$.
Transaction costs set to 2bps per transaction.\label{fig:debt}}
\end{figure}

\begin{figure}
\includegraphics[width=0.4\textwidth]{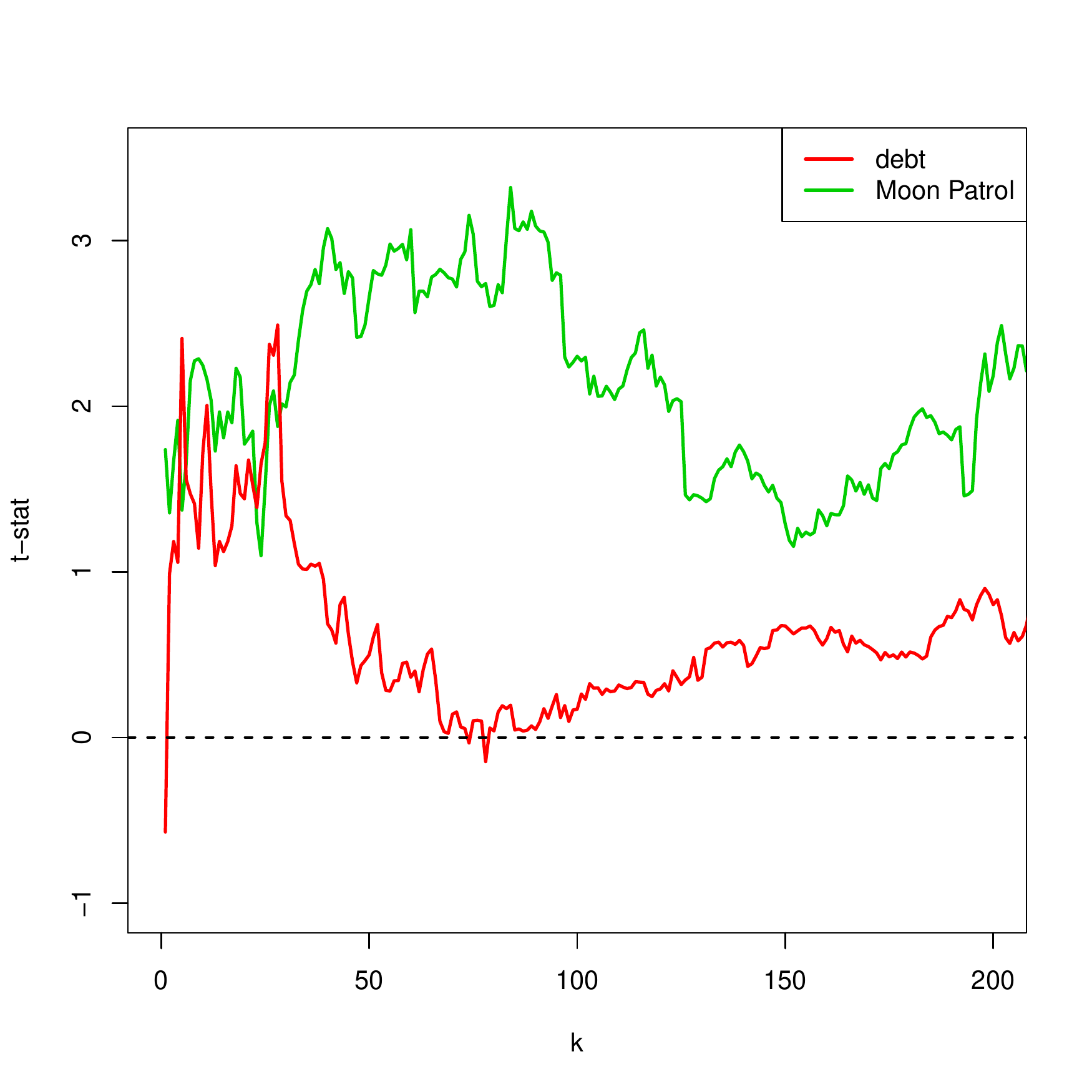}\caption{T-stats of the performance associated with keywords \texttt{debt}
and \texttt{Moon Patrol} versus the length of the moving average $k$.
Transaction costs set to 2bps per transaction.\label{fig:MP-debt-tstat-k}}
\end{figure}

Let us perform some in-sample parameter tuning. The strategy proposed
has only one parameter once the financial asset has been chosen, the
number of time-steps over which the moving average $\bar{v}_{t}$
is performed. Figure \ref{fig:debt} reports the t-tstat of the performance
associated with keyword \texttt{debt} as a function of $k$. Its sign
is relatively robust against changes over the range of $k\in{2,\cdots,30}$
but its typical value in this interval is not particularly exceptional
(between 1 and 2). Let us take now the absolute best keyword from
the four sets, \texttt{Moon Patrol}. Both the values and stability
range of its t-stat are way better than those of \texttt{debt} (see
Figure \ref{fig:MP-debt-tstat-k}), but this is most likely due to
pure chance. There is therefore no reason to trust more one keyword
than the other.

\subsection{No transaction fees}

Assuming an average cost of 2bps (0.02\%) per trade, 104 trades per
year and 8 years of trading (2004-2011), transaction fees diminish
the performance associated to any keyword by about 20\%. As a beneficial
side effect, periods of flat fees-less performance suddenly become
negative performance periods when transaction costs are accounted
for, which provides more realistic expectations. Cost related to spread
and price impact should also included in a proper backtest.

\begin{figure}

\includegraphics[width=0.4\textwidth]{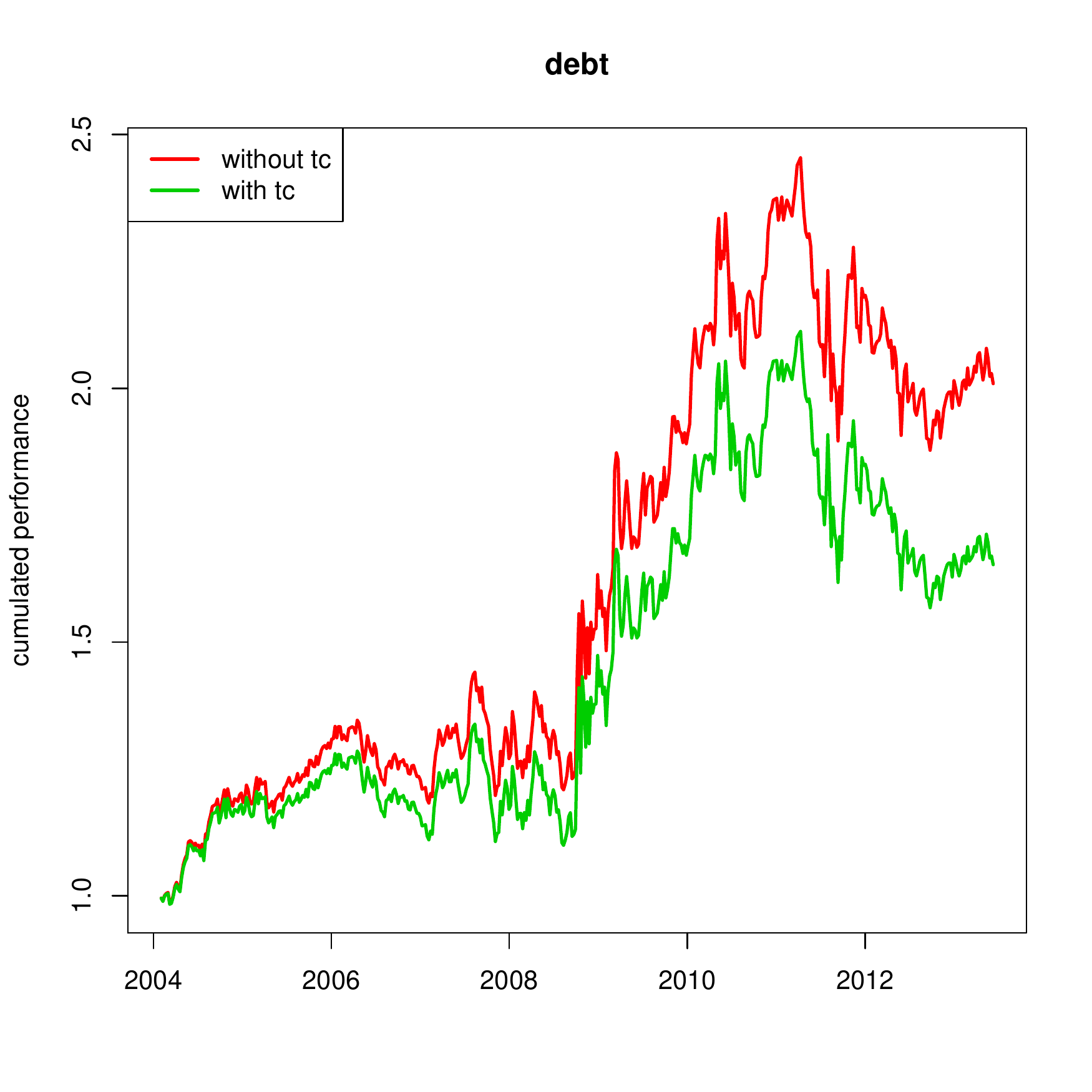}\caption{Cumulative performance associated with keyword \texttt{debt} for $k=3$
with and without transaction costs, set to 2bps.}

\end{figure}

\section{The predictive power of Google Trends}

\begin{figure}
\includegraphics[width=0.4\textwidth]{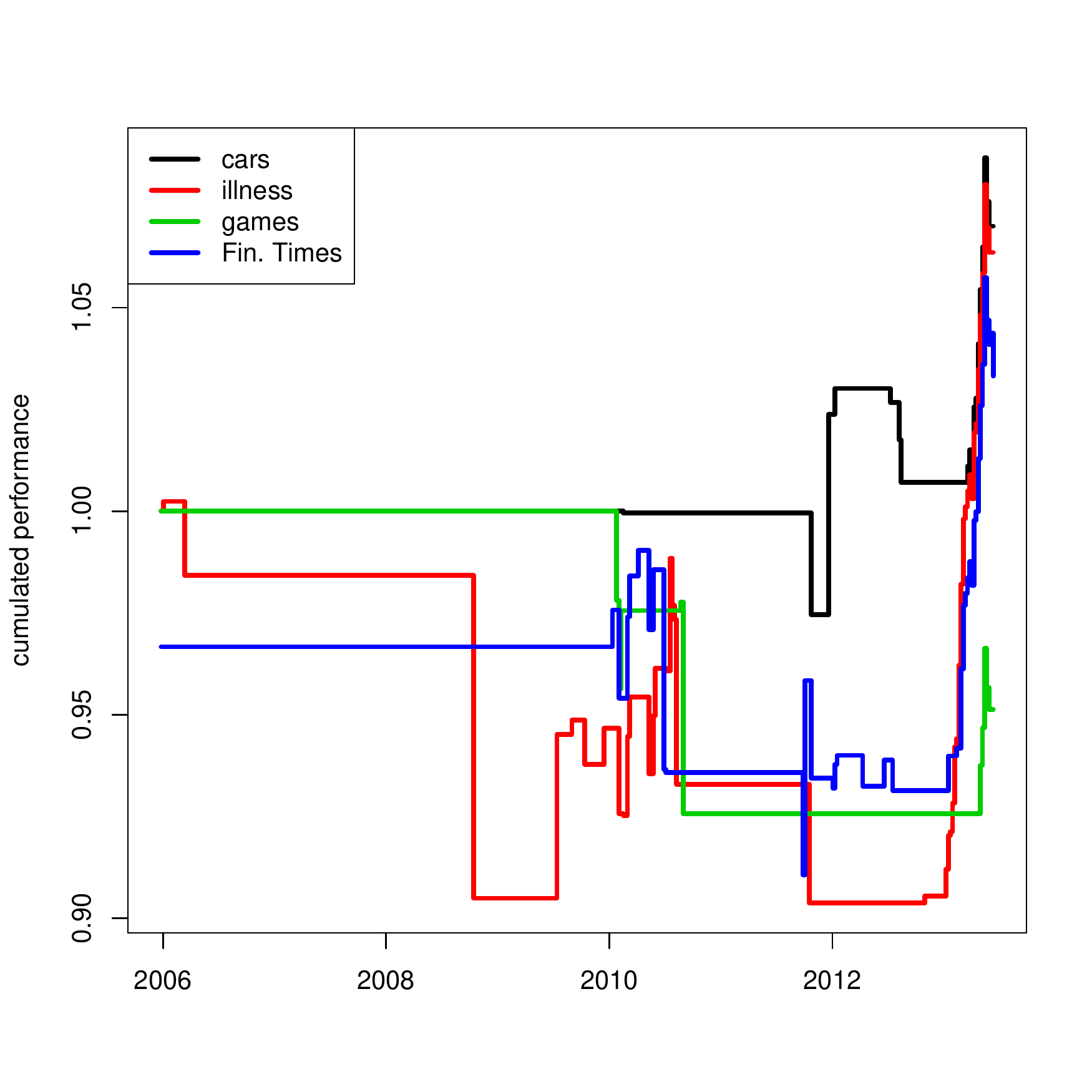}\includegraphics[width=0.4\textwidth]{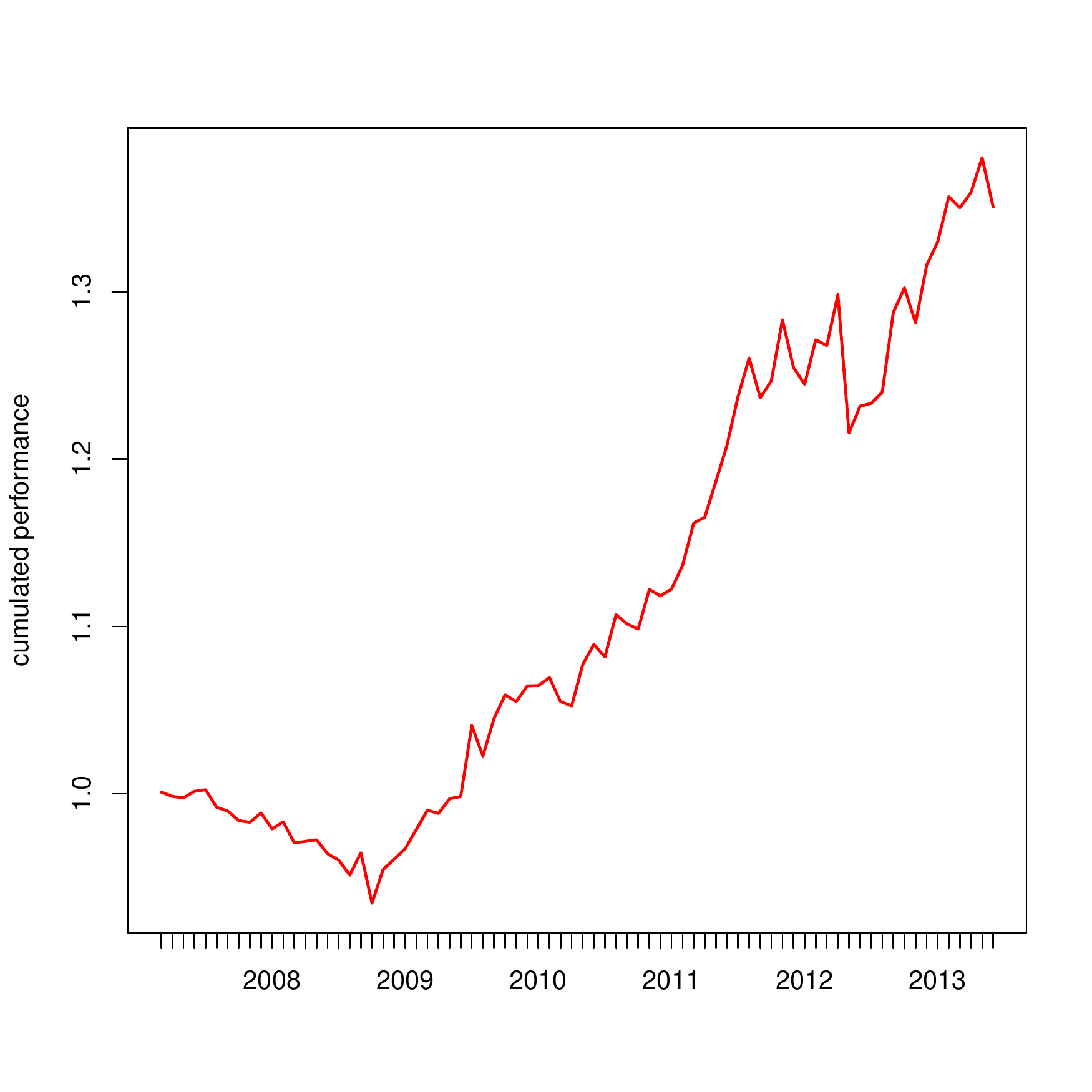}\caption{Left plot: cumulated performance associated with each of the four
keyword sets from 2005-12-23 to 2013-06-14. Right plot: cumulated
performance of suitable keywords applied on suitable assets. Transaction
costs set at 2bps per trade.\label{fig:backtests_4}}
\end{figure}

Given the many methodological weaknesses listed above, one may come
to doubt the conclusions of \cite{preis2013quantifyingGT}. We show
here that they are correct. The first step is to avoid methodological
problems listed above. One of us has used an industrial-grade backtest
system and more sophisticated strategies (which therefore cause tool
bias). First, let us compare the resulting cumulated performance of
the three random keyword sets that we defined, plus the set of keywords
from the Financial Times. For each sets of keywords, we choose as
inputs the raw SVI, lagged SVI, and various moving averages of SVI,
together with past index returns. It turns out that none of the keyword
sets brings information able to predict significantly index movements
(see Fig.~\ref{fig:backtests_4}). This is not incompatible with
results of \cite{da2011search,dzielinski2012measuring,preis2013quantifyingGT}.
It simply means that the signal is probably too weak to be exploitable
in practice. The final part of the performances is of course appealing,
but this come from the fact that Monday close to Friday close SPY
returns have been mostly positive during this period: any machine
learning algorithm applied on returns alone would likely yield the
same result. 

So far we can only conclude that a given proper (and not overly stringent)
backtest system was not able to find any exploitable information from
the four keyword sets, not that the keyword sets do not contain enough
predictive information. To conclude, we use the same backtest system
using some GT data with exactly the same parameters and input types
as before. The resulting preliminary performance, reported in Fig.~\ref{fig:backtests_4},
is more promising and shows that there really is consistently some
predictive information in GT data. It is not particularly impressive
when compared to the performance of SPY itself, but is nevertheless
interesting since the net exposure is always close to zero (see \cite{dc_encelade_2013}
for more information).

\section{Discussion}

Sophisticated methods coupled with careful backtest are needed to
show that Google Trends contains enough exploitable information. This
is because such data include too many searches probably unrelated
to the financial assets for a given keyword, and even more unrelated
to actual trading. When one restricts the searches by providing more
keywords, GT data often only contain information at a monthly time
scale, or no information at all. 

If one goes back to the algorithm proposed by \cite{preis2013quantifyingGT}
and the compatible findings of \cite{da2011search,dzielinski2012measuring},
it is hard to understand why future prices should systematically revert
after a positive SVI surprise and vice-versa one week later. The reversal
is weak and only valid on average. It may be the most frequent outcome,
but profitability is much higher if one knows what triggers reversal
or trend following. There is some evidence that supplementing GT data
with news leads to much improved trading performance (see e.g. \cite{quant3.0mood}).

Another paper by the same group suggests a much more promising source
of information: it links the changes in the number of visits on Wikipedia
pages of given companies to future index returns \cite{moat2013quantifying}.
Further work will investigate the predictive power of this type of
data.

We acknowledge stimulating discussions with Frédéric Abergel, Marouanne
Anane and Thierry Bochud.

\bibliographystyle{plainnat}
\bibliography{biblio}

\appendix

\section{Keywords}

We have downloaded GT data for the following keywords, without any
manual editing.

\subsection{Illnesses}

Source:\href{http://www.ranker.com/list/list-of-common-diseases-most-common-illnesses/diseases-and-medications-info}{http://www.ranker.com/list/list-of-common-diseases-most-common-illnesses/diseases-and-medications-info},
accessed on 27 May 2013\\
\texttt{\footnotesize AIDS, Acne, Acute bronchitis, Allergy, Alopecia,
Altitude sickness, Alzheimer's disease, Andropause, Anorexia nervosa,
Antisocial personality disorder, Arthritis, Asperger syndrome, Asthma,
Attention deficit hyperactivity disorder, Autism, Avoidant personality
disorder, Back pain, Bad Breath, Bedwetting, Benign prostatic hyperplasia,
Bipolar disorder, Bladder cancer, Bleeding, Body dysmorphic disorder,
Bone cancer, Borderline personality disorder, Bovine spongiform encephalopathy,
Brain Cancer, Brain tumor, Breast cancer, Burns, Bursitis, Cancer,
Canker Sores, Carpal tunnel syndrome, Cervical cancer, Cholesterol,
Chronic Childhood Arthritis, Chronic Obstructive Pulmonary Disease,
Coeliac disease, Colorectal cancer, Conjunctivitis, Cradle cap, Crohn's
disease, Dandruff, Deep vein thrombosis, Dehydration, Dependent personality
disorder, Depression, Diabetes mellitus, Diabetes mellitus type 1,
Diaper rash, Diarrhea, Disabilities, Dissociative identity disorder,
Diverticulitis, Down syndrome, Drug abuse, Dysfunctional uterine bleeding,
Dyslexia, Ear Infections, Ear Problems, Eating Disorders, Eczema,
Edwards syndrome, Endometriosis, Epilepsy, Erectile dysfunction, Eye
Problems, Fibromyalgia, Flu, Fracture, Freckle, Gallbladder Diseases,
Gallstone, Gastroesophageal reflux disease, Generalized Anxiety Disorder,
Genital wart, Glomerulonephritis, Gonorrhoea, Gout, Gum Diseases,
Gynecomastia, HIV, Head Lice, Headache, Hearing impairment, Heart
Disease, Heart failure, Heartburn, Heat Stroke, Heel Pain, Hemorrhoid,
Hepatitis, Herniated Discs, Herpes simplex, Hiatus hernia, Histrionic
personality disorder, Hyperglycemia, Hyperkalemia, Hypertension, Hyperthyroidism,
Hypothyroidism, Infectious Diseases, Infectious mononucleosis, Infertility,
Influenza, Iron deficiency anemia, Irritable Male Syndrome, Irritable
bowel syndrome, Itching, Joint Pain, Juvenile Diabetes, Kidney Disease,
Kidney stone, Leukemia, Liver tumour, Lung cancer, Malaria, Melena,
Memory Loss, Menopause, Mesothelioma, Migraine, Miscarriage, Mucus
In Stool, Multiple sclerosis, Muscle Cramps, Muscle Fatigue, Muscle
Pain, Myocardial infarction, Nail Biting, Narcissistic personality
disorder, Neck Pain, Obesity, Obsessive-compulsive disorder, Osteoarthritis,
Osteomyelitis, Osteoporosis, Ovarian cancer, Pain, Panic attack, Paranoid
personality disorder, Parkinson's disease, Penis Enlargement, Peptic
ulcer, Peripheral artery occlusive disease, Personality disorder,
Pervasive developmental disorder, Peyronie's disease, Phobia, Pneumonia,
Poliomyelitis, Polycystic ovary syndrome, Post-nasal drip, Post-traumatic
stress disorder, Premature birth, Premenstrual syndrome, Propecia,
Prostate cancer, Psoriasis, Reactive attachment disorder, Renal failure,
Restless legs syndrome, Rheumatic fever, Rheumatoid arthritis, Rosacea,
Rotator Cuff, Scabies, Scars, Schizoid personality disorder, Schizophrenia,
Sciatica, Severe acute respiratory syndrome, Sexually transmitted
disease, Sinusitis, Skin Eruptions, Skin cancer, Sleep disorder, Smallpox,
Snoring, Social anxiety disorder, Staph infection, Stomach cancer,
Strep throat, Sudden infant death syndrome, Sunburn, Syphilis, Systemic
lupus erythematosus, Tennis elbow, Termination Of Pregnancy, Testicular
cancer, Tinea, Tooth Decay, Traumatic brain injury, Tuberculosis,
Ulcers, Urinary tract infection, Urticaria, Varicose veins.}{\footnotesize \par}

\subsection{Classic cars}

Source:\href{http://www.ranker.com/crowdranked-list/the-best-1960_s-cars}{http://www.ranker.com/crowdranked-list/the-best-1960\_{}s-cars},
accessed on 27 May 2013\\
\texttt{\footnotesize 1960 Aston Martin DB4 Zagato, 1960 Ford, 1961
Ferrari 250 SWB, 1961 Ferrari 250GT California, 1963 Corvette, 1963
Iso Griffo A3L, 1964 Ferrari 250 GTL (Lusso), 1965 Bizzarrini 5300
Strada, 1965 Ford GT40, 1965 Maserati Mistral, 1965 Shelby Cobra,
1966 Ferrari 365P, 1966 Maserati Ghibli, 1967 Alfa Romeo Stradale,
1967 Ferrari 275 GTB/4, 1967 Shelby Mustang KR500, 1968 Chevrolet
Corvette L88, 1968 DeTomaso Mangusta, 1969 Pontiac Trans Am, 1969
Yenko Chevelle, 57 Chevy, 68 Ferrari 365 GTB/4Daytona Spyder, 69 Yenko
Camaro Z28, AC Cobra, Alfa Romeo Spider, Aston Martin DB5, Austin
Mini Saloon 1959, BMW E9, Buick Riviera, Buick Wildcat, Cane, Chevrolet
Camaro, Chevrolet Chevelle, Chevrolet Impala, Chevy Chevelle, Chrysler
Valiant, Corvette Stingray, Dodge Challenger, Dodge Charger, Dodge
Dart Swinger, Facel Vega Facel II, Ferrari 250, Ferrari 250 GTO, Ferrari
250 GTO, Ferrari 275, Ferrari Daytona, Fiat 500, Ford Corsair, Ford
Cortina, Ford GT40, Ford Mustang, Ford Ranchero, Ford Thunderbird,
Ford Torino, Ford Zephyr MK III, Iso Grifo, Jaguar E-type, Jeep CJ,
Lamborghini Miura, Lamborghini Miura SV, Lincoln Continental, Lotus
Elan, Maserati Ghibli, Mercedes Benz 220SE, Mercedes-Benz 300SL, Mercury
Cougar, Plymouth Barracuda, Pontiac GTO, Porsche 356, Porsche 911,
Porsche 911, Porsche 911 classic, Rambler Classic, Rover 2000, Shelby
Daytona Coupe, Shelby GT350, Shelby GT500, Studebaker Avanti, Sunbeam
Tiger, Toyota 2000GT, Triumph 2000, Vauxhall Velox 1960, Vauxhall
Victor 1963, Wolseley 15/60}{\footnotesize \par}

\subsection{Arcade Games}

Source:\href{"http://www.ranker.com/list/list-of-common-diseases-most-common-illnesses/diseases-and-medications-info}{http://www.ranker.com/list/list-of-common-diseases-most-common-illnesses/diseases-and-medications-info},
accessed on 27 May 2013\\
\texttt{\footnotesize 1942, 1943, 720\textdegree{}, After Burner,
Airwolf, Altered Beast, Arkanoid, Asteroids, Bad Dudes Vs. DragonNinja,
Bagman, Battlezone, Beamrider, Berzerk, Bionic Commando, Bomb Jack,
Breakout, Bubble Bobble, Bubbles, BurgerTime, Centipede, Circus Charlie,
Commando, Crystal Castles, Cyberball, Dangar - Ufo Robo, Defender,
Dig Dug, Donkey Kong, Donkey Kong 3, Donkey Kong Junior, Double Dragon,
Dragon's Lair, E.T. (Atari 2600), Elevator Action, Final Fight, Flashback,
Food Fight, Frogger, Front Line, Galaga, Galaxian, Gauntlet, Geometry
Wars, Gorf, Gorf, Gyruss, Hogan's Alley, Ikari Warriors, Joust, Kangaroo,
Karate Champ, Kid Icarus, Lode Runner, Lunar Lander, Manic Miner,
Mappy, Marble Madness, Mario Bros., Millipede, Miner 2049er, Missile
Command, Moon Buggy, Moon Patrol, Ms. Pac-Man, Naughty Boy, Pac-Man,
Paperboy, Pengo, Pitfall!, Pole Position, Pong, Popeye, Punch-Out!!,
Q{*}bert, Rampage, Red Baron, Robotron: 2084, Rygar: The Legendary
Adventure, Sewer Sam, Snow Bros, Space Invaders, Spy Hunter, Star
Wars, Stargate, Street Fighter, Super Pac-Man, Tempest, Tetris, The
Adventures of Robby Roto!, The Simpsons, Time Pilot, ToeJam \& Earl,
Toki, Track \& Field, Tron, Wizard Of Wor, Xevious }{\footnotesize \par}

\section{Source code}

Here is a simple implementation in R of the strategy given in \cite{preis2013quantifyingGT}.
We do mean ``\texttt{=}'' instead of ``\texttt{<-}''.

\begin{lstlisting}
computePerfStats=function(filename,k=10,getPerf=FALSE){
	gtdata=loadGTdata(filename)
	if(is.null(gtdata) || length(gtdata)<100){
		return(NULL)
	} 
	spy=loadYahooData('SPY')                
	spy_rets=getFutureReturns(spy)          #spy_rets is a zoo object, contains r_{t+1}
	gtdata_mean=rollmeanr(gtdata,k)         # \bar v_t
	gtdata_mean_lagged=lag(gtdata_mean,-1)  # \bar v_{t-1}
	pos=2*(gtdata>gtdata_mean_lagged)-1      
	perf=-pos*spy_rets
	perf=perf[which(!is.na(perf))]
	if(getPerf){
		return(perf)
	}else{
		return(t.test(perf)$statistic)
	}
} 

\end{lstlisting}

\end{document}